\documentclass[aps,prb,twocolumn,10pt,floatfix]{revtex4-2}

\usepackage{amssymb}
\usepackage{amsfonts}
\usepackage{amsmath}
\usepackage[pdftex]{graphicx} 
\usepackage{microtype}
\usepackage{grffile}
\usepackage{braket}
\usepackage{mdframed}
\usepackage[unicode=true,pdfusetitle,bookmarks=true,bookmarksnumbered=false,bookmarksopen=false,breaklinks=false,pdfborder={0 0 1},backref=false,colorlinks=true,citecolor=blue,urlcolor=blue,linkcolor=blue]{hyperref}
\usepackage{physics}
\usepackage{inputenc}
\usepackage{dcolumn}
\usepackage{parskip}
\usepackage[usenames]{xcolor}

\usepackage{natbib}

\renewcommand{\abs}[1]{\left| #1  \right| }
\newcommand{\avg}[1]{\left< #1  \right> }

\newcommand{\ZZ}{\mathbb{Z}}

\newmdenv[backgroundcolor=gray!40,hidealllines=true]{infobox}

\begin{document}

\title{Lifetime of Excitations in Atomic and Molecular Bose-Einstein Condensates}

\author{Matteo Bellitti}
\affiliation{Department of Physics, Boston University, Boston, MA 02215, USA}
\author{Garry Goldstein}
\affiliation{Department of Physics, Boston University, Boston, MA 02215, USA}
\author{Chris R. Laumann}
\affiliation{Department of Physics, Boston University, Boston, MA 02215, USA}

\date{\today}

\begin{abstract}
Recent experimental progress has produced Molecular Superfluids (MSF) in thermal equilibrium; this opens the door to a new class of experiments investigating the associated thermodynamic and dynamical responses. 
We review the theoretical picture of the phase diagram and quasiparticle spectrum in the Atomic Superfluid (ASF) and MSF phases. 
We further compute the  parametric dependence of the quasiparticle lifetimes at one-loop order.
In the MSF phase, the $U(1)$ particle number symmetry breaks to $\ZZ_2$ and the spectrum exhibits a gapless Goldstone mode in addition to a gapped $\ZZ_2$-protected atom-like mode.
In the ASF phase, the $U(1)$ symmetry breaks completely, leaving behind a Goldstone mode and an unprotected gapped mode.
In both phases, the Goldstone mode decays with a rate given by the celebrated Belyaev result, as in a single component condensate.
In the MSF phase, the gapped mode is sharp up to a critical Cherenkov momentum beyond which it emits phonons.
In the ASF phase, the gapped mode decays with a constant rate even at small momenta.
These decay rates govern the spectral response in microtrap tunneling experiments and lead to sharp features in the transmission spectrum of atoms fired through molecular clouds.
\end{abstract}

\maketitle


\section{Introduction} 
\label{sec:introduction}

\begin{figure*}
  \centering
  \includegraphics[width=\linewidth]{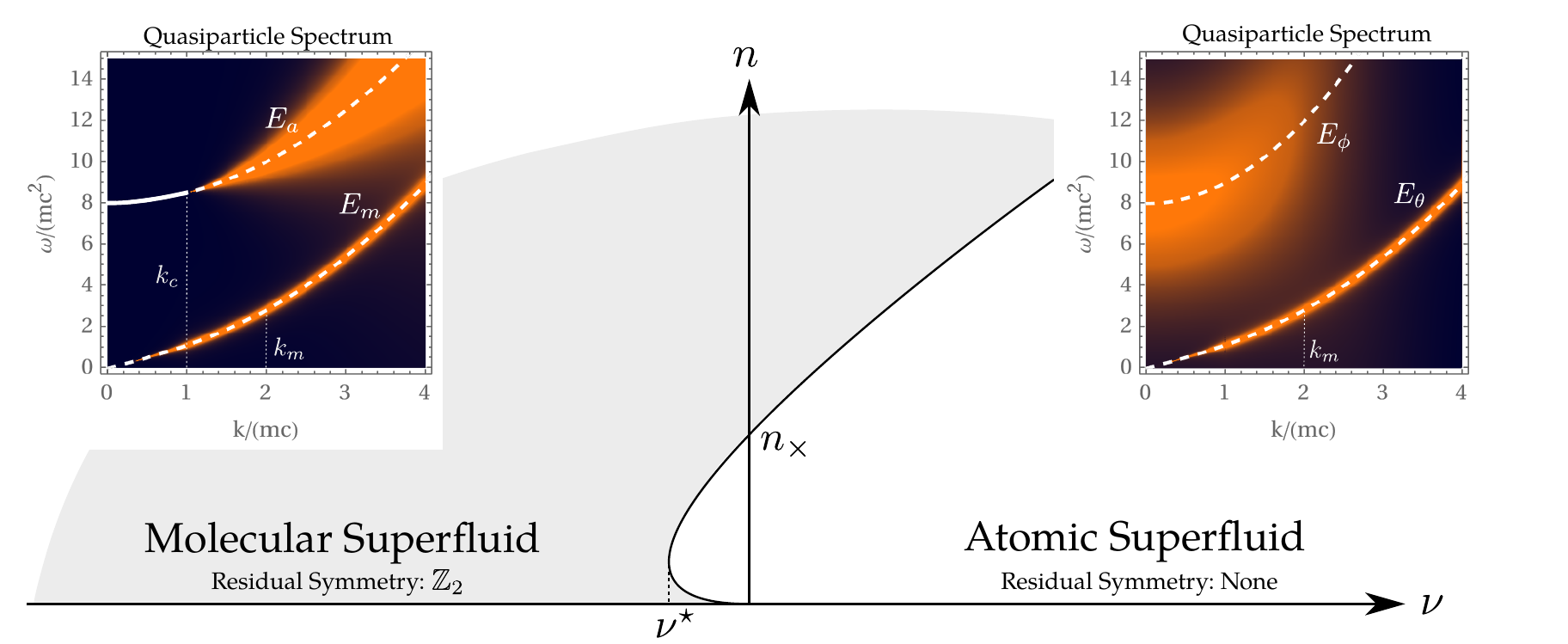}
  \caption{The mean field phase diagram as a function of the binding energy $\nu$ and total condensate density $n = n_a + 2 n_m$, with two insets showing the quasiparticle spectral weight. The white lines are the mean field results for the quasiparticle dispersion discussed in Sec.\ref{sec:bogoliubov_theory_of_the_molecular_superfluid}. The solid white line in the MSF side inset indicates that the excitation is infinitely long lived:
  in the MSF phase the $\mathbb{Z}_2$ symmetry keeps the spectral line for the gapped excitation sharp up to a threshold momentum $k_c$, above which decay by phonon emission is allowed. In the ASF phase there is no such symmetry protection, the gapped mode is damped at arbitrarily low momenta and is in fact very diffuse. In both phases the gapless mode is always damped. To give the reader a reference for the momentum scales we marked $k_m$ on the $k$ axis: it is the scale where the gapless excitation changes nature from predominantly phonon--like to particle--like. The phase boundary depicted assumes $2 g_{am} > g_m$, see Eq.\eqref{eq:phase_boundary_eq}.}
  \label{fig:phase_diagram_finitealpha}
\end{figure*}

The study of the Molecular Superfluid (MSF) phase of weakly interacting ultracold Bose atoms, first analyzed theoretically some two decades ago \cite{radzihovsky2004,radzihovsky2008}, has recently heated up again \cite{zhang2022,malla2022,lam2022} due to groundbreaking  experimental progress \cite{zhang2021a} in coherently trapping ultracold cesium atoms and controlling their Feshbach resonances to produce cesium molecules (Cs$_2$) \cite{chin2010,berninger2013,mark2007,chin2003,kohler2006}.
%
The refinement of these trapping techniques enables a new generation of experiments probing both equilibrium and dynamical properties of both the MSF and proximate Atomic Superfluid (ASF) phase. 
While the thermodynamics of the MSF-ASF system are well-known theoretically \cite{radzihovsky2008,romans2004} (see Fig.~\ref{fig:phase_diagram_finitealpha} for a phase diagram), its dynamical responses are more complicated. 
A full dynamical theory of the system requires an understanding of both the quasiparticle content and their dissipative scattering properties.
   
In this work, we compute the near equilibrium decay rates of the quasiparticles in the ASF and MSF phases to one-loop order at zero temperature. 
The quasiparticle decay rate determines the width of the spectral function, schematically illustrated in the insets of Fig.~\ref{fig:phase_diagram_finitealpha}. 
In principle, this is directly measurable by tunneling experiments in which a micro-trap is placed in tunnel contact with the MSF \cite{micheli2004}.

Single atom transmission spectroscopy provides an alternative, and perhaps more striking, experimental signature of the quasiparticle dynamics in the MSF phase. 
An incident atom evolves into the gapped quasiparticle mode, which is sharp up to a critical momentum $k_c$.   
Beyond this threshold, the quasiparticle Cherenkov radiates phonons (the gapless mode). 
Thus, a slow atom fired through a MSF cloud propagates without dissipation, while faster atoms slow down until they drop below the Cherenkov threshold. 
This leads to a sharp feature in the energy spectrum of the transmitted atoms as the incident energy crosses the threshold  -- assuming the cloud is ``optically dense'' enough to slow the atom before it passes through.
Using our computed scattering rates, we estimate that the stopping power of a typical molecular cloud is sufficient to observe these features (see Fig.~\ref{fig:decay_length}).
The existence of $k_c$ follows from the symmetry structure of the MSF phase, as we discuss below. Deep in the MSF phase $k_c \simeq m c$, where $m$ is the atomic mass and $c$ the speed of sound (see Eq.\eqref{eq:k_c}).

We summarize here the equilibrium properties of the system to contextualize our work and keep the presentation self--contained.
The system has a global $U\left(1\right)$ symmetry associated with total atom number conservation $n=N_{a}+2N_{m}$, where $N_{a}$ is the number of atoms and $N_{m}$ is the number of molecules. 
The simplest model Hamiltonian for the system \cite{tommasini1998} includes kinetic contributions, density--density interactions, and a Feshbach interaction that coherently converts two atoms into a molecule and vice versa - see equations \eqref{eq:euclidean_action} through \eqref{eq:feshbach_term}. 
%
This interconversion occurs thanks to the hyperfine interactions between the closed and open scattering channels for atomic collisions \cite{timmermans1999,chin2010,duine2004}.

With the formation of the condensate the global $U\left(1\right)$ is spontaneously broken. At low temperature there are two distinct scenarios for this $U\left(1\right)$ breaking: 1. when just the molecules condense the phase is known as molecular superfluid (MSF), and 2. when both the atoms and the molecules condense as atomic superfluid (ASF). Due to the coherent interconversion process, the condensation of the atoms forces a condensation of the molecules, and thus there is no phase where the atoms are condensed but the molecules are not. In the MSF phase the global $U\left(1\right)$ is reduced to a global $\mathbb{Z}_{2}$, where the $\mathbb{Z}_2$ charge is the parity of the atom number, while
in the ASF phase there is no remaining symmetry. 
Prior work has argued \cite{lee2004,radzihovsky2004} that the zero temperature quantum phase transition between the MSF and ASF phases is continuous and lies in the quantum Ising class. 

There are two experimentally tunable parameters which control the phase diagram: the total number of atoms $n$ and the molecular binding energy $\nu$. 
When $\nu$ is large and positive, the molecular state is anti--bound, so we expect ASF to be the equilibrium phase, while for large and negative $\nu$ forming a molecule is energetically favorable and we expect MSF. 
No such simple argument can be made for the effect of the particle number $n$ on the phase.
Indeed, the transition is re-entrant as a function of particle number at the mean--field level (see Fig. \ref{fig:phase_diagram_finitealpha}). 


The quasiparticle content of the ASF and MSF phases is as follows.
Since the $U\left(1\right)$ symmetry is broken in both the ASF and MSF phases, there is one Goldstone boson on each side of the transition. 
In the MSF phase, the molecules are at lower energy than the atoms and the Goldstone mode is molecular in nature, carrying a $\mathbb{Z}_{2}$ even charge. 
A second, gapped branch of the spectrum is atom-like, carrying odd $\ZZ_2$ charge. 
By $\mathbb{Z}_{2}$ conservation, the upper mode can lose energy and momentum into the Goldstone mode, but cannot disappear.
At zero temperature, such emission can only happen above a Cherenkov momentum $k_{c}$, where the group velocity matches the Goldstone velocity. 
The decay width above the Cherenkov transition is proportional to $\sim\left(k-k_{c}\right)^{3}$ and is given by Eq.\eqref{eq:gamma_msf}. 
The gapless mode in the MSF phase has a similar lifetime to that of a single component BEC \cite{beliaev1958} and is given by Eq.\eqref{eq:gamma_single_component}. 

In the ASF phase the Goldstone mode has mostly atomic character while the high energy mode has mostly molecular character.
However, there is no $\mathbb{Z}_{2}$ conservation so no decays are forbidden. The decay rate of the high energy mode scales with the gap, $\Delta$, as $\Delta^{4}$ and is given by Eq.\eqref{eq:gamma_asf}. 
The decay rate of the Goldstone mode in the ASF phase is of similar form to Eq.\eqref{eq:gamma_single_component} however with $2m\rightarrow m$. 

To obtain these results, we derive a low energy effective theory on each side of the transition from the  microscopic theory of Eq.\eqref{eq:euclidean_action}. 
This approach significantly simplifies the calculation compared to using the microscopic theory directly, and is applicable as long as we restrict our attention to excitations with wavelengths longer than the healing length of the condensate and energies below the multiparticle threshold. 
Using the effective theory, we compute for each excitation the one--loop imaginary part of the self energy in the on--shell approximation, and thus estimate the decay rates of the gapped modes.

The paper is organized as follows: in Sec.\ref{sec:hamiltonian} we set up the problem, then in Sec.\ref{sec:phase_diagram} we briefly review the mean--field phase diagram. In Sec.\ref{sec:bogoliubov_theory_of_the_molecular_superfluid} we review the Bogoliubov mean field theory of the spectrum and in Sec.\ref{sec:lifetimes} we summarize our results for the decay rates of the gapped mode in each phase. The interested reader will find details of the calculation of certain integrals in Appendix \ref{sub:off_shell_cherenkov}.

\section{Euclidean Action and Symmetries}\label{hamiltonian}
\label{sec:hamiltonian}
The number of atoms $N_a$ and the number of molecules $N_m$ are not separately conserved, only the total number $N = N_a + 2N_m$. The Euclidean action of the system is  
\begin{align}
\label{eq:euclidean_action}
S &= 
\int d\mathbf{r} d \tau \,
\big(
\mathcal{T} + \mathcal{H}_a + \mathcal{H}_m + \mathcal{H}_{am} + \mathcal{H}_F 
\big)
\\[2mm]
\mathcal{T} &= \bar{\Psi}_m \partial_\tau \Psi_m + \bar{\Psi}_a \partial_\tau \Psi_a 
\\[2mm]
\mathcal{H}_m &= 
\bar{\Psi}_m
\bigg(-\frac{\nabla^2}{4 m} - 2\mu + \nu\bigg)
{\Psi}_m
+
\frac{g_m}{2}
\abs{{\Psi}_m}^4
\\[2mm]
\mathcal{H}_a &= 
\bar{\Psi}_a
\bigg(-\frac{\nabla^2}{2 m} -\mu\bigg)
{\Psi}_a
+
\frac{g_a}{2}
\abs{{\Psi}_a}^4
\\[2mm]
\mathcal{H}_{am} &= 
g_{am}
\abs{{\Psi}_a}^2
\abs{{\Psi}_m}^2
\\[2mm]
\mathcal{H}_F &= -
\alpha
\big(
\bar{\Psi}_a^2
\Psi_m
+
\bar{\Psi}_m
\Psi_a^2
\big) \label{eq:feshbach_term}
\end{align}
The subscript $a$ denotes the atoms and $m$ the molecules. Here $\nu$ is the molecular binding energy; a negative $\nu$ means it is energetically favorable to make bound states. The minus sign in front of the Feshbach term is chosen so that $\alpha > 0$ in equilibrium the phases of the condensates are locked to the same value; Without loss of generality we assume $\alpha > 0$, as it is always possible to absorb its sign with the field redefinition ${\Psi}_m \to -{\Psi}_m$. 
The chemical potential $\mu$ appears with a factor of $2$ in the molecular part because two atoms bind to form a molecule and only the total number is conserved. Under the $U(1)$ symmetry associated to this conservation law the atomic field transforms with unit charge and the molecular one with double charge
\begin{align}
\Psi_a \to e^{i \theta} \Psi_a \qquad
\Psi_m \to e^{i 2 \theta} \Psi_m \qquad
\end{align}
The system has two nontrivial low--temperature phases: the atomic superfluid (ASF) phase, in which the $U(1)$ symmetry is completely broken --both $\Psi_a$ and $\Psi_m$ condense--, and
the molecular superfluid (MSF) phase, in which the $U(1)$ breaks down to $\mathbb{Z}_2$
\begin{align}\Psi_a \to -\Psi_a \qquad \Psi_m \to \Psi_m\end{align}
As such we expect any continuous transition between the two phases to be Ising class \cite{radzihovsky2008}.

\hypertarget{mean-field-phase-diagram}{%
\section{Mean Field phase diagram}\label{mean-field-phase-diagram}}
\label{sec:phase_diagram}
We briefly review the mean field phase diagram (see Fig.\ref{fig:phase_diagram_finitealpha}) of the system, to provide context for our calculations. We map the phase diagram in terms of the total density $n = n_a + 2 n_m$ and the binding energy $\nu$, which are experimentally tunable. As anticipated, there are two stable thermodynamic phases at zero temperature (ASF and MSF), separated by an Ising class transition line \cite{radzihovsky2004,lee2004}. If $2 g_{am} > g_m$ the transition is reentrant as a function of $n$ for small negative $\nu$ (see Eq.\eqref{eq:phase_boundary_eq}). 
%
%
Parametrizing the fields in polar form
\begin{align}
\Psi_{j} = \sqrt{n_j} e^{i\theta_j}
\qquad
j = a,m
\end{align}
clarifies the role of the relative phase between the condensates. In the absence of an external potential the equilibrium solution is uniform, so we drop the gradient terms. The mean field energy density is then
\begin{align}
\label{eq:mf_energy}
\begin{split}
\mathcal{E}_{MF} =
\frac{1}{2} g_a n_a^2 +
\frac{1}{2} g_m n_m^2 +
g_{am} n_a n_m 
\\
- 2 \alpha n_a \sqrt{n_m} \cos{(2 \theta_a - \theta_m)}
\\
- \mu (n_a + 2 n_m - n)
+ \nu n_m
\end{split}
\end{align}
which is minimized by 
\begin{align}
\label{eq:phase_lock_eq}
0 &= \alpha n_a \sqrt{n_m} \sin(2 \theta_a - \theta_m)
\\
\mu &= g_a n_a + g_{am} n_m - 2 \alpha \sqrt{n_m} \cos(2\theta_a - \theta_m) 
\label{eq:mu_equil_1}
\\
2 \mu - \nu &= g_m n_m + g_{am} n_a - \alpha \frac{n_a}{\sqrt{n_m}} \cos(2\theta_a - \theta_m) 
\label{eq:mu_equil_2}
\end{align}
These equations have two nontrivial solutions:
\begin{description}
    \item[MSF phase] $n_a = 0, n_m = n/2 > 0$. At mean field 
    \begin{align}
    2 \mu - \nu = g_m n_m
    \end{align}
    as usual for a weakly interacting Bose gas. The combination $2 \mu - \nu$ acts as an effective chemical potential for the molecules.
    \item[ASF phase] $n_a > 0, n_m > 0$. While there is a closed form expression for $n_a$ and $n_m$, it is complicated and we omit it. In the large $\nu$ limit at fixed density $n$, we find to lowest order in $\nu$ the relations $\sqrt{n_m} = \alpha n/\nu$ and $\mu = g_a n$, which means that the system behaves as a simple atomic BEC.
\end{description}
There is no phase where the atoms are condensed but the molecules are not. 
Since $\alpha > 0$, in the ASF phase the phases of the atomic and molecular condensates lock together:
\begin{align}
   2 \theta_a - \theta_m = 0 
\end{align}
Eliminating the chemical potential and imposing $n_a = 0$ we find the phase boundary (see Fig.\ref{fig:phase_diagram_finitealpha})
\begin{align}
\label{eq:phase_boundary_eq}
\bigg(g_{am} - \frac{g_m}{2}\bigg) n - 2\alpha \sqrt{2n} = \nu
\end{align} 
The phase boundary crosses the $\nu = 0$ axis at
\begin{align}
n_\times = 32 \alpha^2/(2g_{am} - g_m)^2
\end{align}
and the leftmost point is at 
\begin{align}
|\nu^\star|= 4 \alpha^2/|2 g_{am} - g_m|
\end{align}
so that using a negative binding energy with $|\nu| < |\nu^\star|$ the reentrant nature of the transition is visible. Finally, we point out that the phase diagram and the order of the transition are sensitive to the sign of $g_a g_m - g_{am}^2$ (for concreteness we assume $g_a g_m > g_{am}^2$) but the realized phases are the same. For a detailed discussion see \cite{radzihovsky2008}.

\section{Bogoliubov theory of the Molecular Superfluid}
\label{sec:bogoliubov_theory_of_the_molecular_superfluid}
In this section we review the Bogoliubov theory of the molecular superfluid \cite{radzihovsky2004}, which describes the quadratic fluctuations around the mean field solutions of Sec.\ref{sec:phase_diagram}. See Ref~.\cite{popov2001} for a detailed review of the Bogoliubov approach. We expand the action around the mean field configuration 
\begin{align}
\avg{\Psi_j} = \sqrt{n_j^0}
\qquad 
j = a,m
\end{align}
where the two constants $(n_a^0,n_m^0)$ solve Eq.\eqref{eq:phase_lock_eq}-\eqref{eq:mu_equil_2}. We redefine the fields to isolate the deviations $\psi_j$ from equilibrium
\begin{align}
\Psi_j = \sqrt{n_j^0} + \psi_j \qquad j = a,m
\end{align}
and we drop from the action all terms of degree higher than two in $\psi_j$. The result is
\begin{align}
S = S_2 + \int d\bm{r} d\tau \, \mathcal{E}_{MF}
\end{align} 
where $\mathcal{E}_{MF}$ is the mean field energy density of Eq.\eqref{eq:mf_energy}, there are no terms linear in the fluctuation fields, and $S_2$ is the quadratic action associated to the fluctuations: 
\begin{widetext}
\begin{align}
\begin{split}
S_2 = \int d\mathbf{r} d\tau
\bigg[
&\bar{\psi}_m \partial_\tau \psi_m
+
\bar{\psi}_a \partial_\tau \psi_a
\\
&+
\bar{\psi}_m
\bigg(
-\frac{\nabla^2}{4m} - 2\mu + \nu
\bigg)
\psi_m
+ \frac{1}{2}
g_m n_m^0
\big(
\psi_m^2
+
\bar{\psi}_m^2 
+ 4  \abs{\psi_m}^2
\big)
\\
&+
\bar{\psi}_a
\bigg(
- \frac{\nabla^2}{2m} - \mu
\bigg)
\psi_a
+ \frac{1}{2}
g_a n_a^0
\big(
\psi_a^2
+
\bar{\psi}_a^2 
+ 4  \abs{\psi_a}^2
\big)
\\
&+
g_{am}
\sqrt{n_a^0 n_m^0}
\big(
\bar{\psi}_m \bar{\psi}_a  
+
\bar{\psi}_m {\psi}_a  
+{\psi}_m \bar{\psi}_a
+
{\psi}_m {\psi}_a
\big)
+
g_{am} 
\big(
n_m^0 |\psi_a|^2 
+
n_a^0 |\psi_m|^2
\big)
\\
&- \alpha
\big(
2 \sqrt{n_a^0} \bar{\psi}_m \psi_a
+
\sqrt{n_m^0} \psi_a^2
+ 
2 \sqrt{n_a^0} {\psi_m} \bar{\psi}_a
+
\sqrt{n_m^0} \bar{\psi}_a^2
\big)
\bigg]
\end{split}
\end{align}
\end{widetext}
It is convenient to work in momentum space and organize the fields in a Nambu spinor
\begin{align}
\Phi_\mathbf{k}^\dagger \equiv
\begin{pmatrix}
\bar{\psi}_{a,\mathbf{k}} &
{\psi}_{a,-\mathbf{k}} &
\bar{\psi}_{m,\mathbf{k}} &
{\psi}_{m,-\mathbf{k}}
\end{pmatrix}
\end{align}
We have already extracted the spatially uniform part from the fields, so the Fourier expansion of $\psi_j$ does not have a $k=0$ term. To avoid double counting the anomalous terms we organize the momentum sum as
%
\begin{align}
S_2 = \sum_n \sum_{\mathbf{k},k_z>0}
\Phi_\mathbf{k}^\dagger
\cdot
G_0^{-1}
\cdot
\Phi_\mathbf{k}
\end{align}
where the inverse $G_0^{-1}$ of the matrix propagator is given in Eq.~\eqref{eq:bogoliubov_propagator_matrix}. Notice that in the MSF phase $n_a^0 = 0$, so that the matrix is block diagonal. This is a manifestation of $\mathbb{Z}_2$ symmetry: the atom--like Bogoliubov quasiparticles only have overlap with 
$\psi_{a,\mathbf{k}}$ and $\bar{\psi}_{a,-\mathbf{k}}$, without a molecular component. The molecular block describes a simple weakly interacting condensate of molecules, while the dynamics of the atomic fluctuations is nontrivial: the atoms are not condensed, but the anomalous averages $\avg{\psi_a \psi_a
}$ acquire a finite value thanks to the off--diagonal Feshbach terms $-2 \alpha \sqrt{n_m^0}$ that provide coherent conversion of a condensed molecule into two atoms.
\begin{widetext}
\begin{align}
\label{eq:bogoliubov_propagator_matrix}
G_0^{-1} = 
\setlength{\arraycolsep}{-5pt}
\begin{pmatrix}
\hspace*{1mm} \omega_n + \xi_a + 2g_a n_a^0 + g_{am} n_m^0 &
g_a n_a^0 - 2 \alpha \sqrt{n^0_m} &
g_{am} \sqrt{n_m^0 n_a^0} - 2 \alpha \sqrt{n_a^0} &
g_{am} \sqrt{n_m^0 n_a^0} 
\\[1em]
g_a n_a^0 - 2 \alpha \sqrt{n_m^0} &
- \omega_n + \xi_a + 2g_a n_a^0 + g_{am} n_m^0  &
g_{am} \sqrt{n_m^0 n_a^0} &
g_{am} \sqrt{n_m^0 n_a^0} - 2 \alpha \sqrt{n_a^0}
\\[1em]
g_{am} \sqrt{n_m^0 n_a^0}- 2 \alpha \sqrt{n_a^0} &
g_{am} \sqrt{n_m^0 n_a^0} &
\omega_n + \xi_m + 2g_m n_m^0 + g_{am} n_a^0 &
g_m n_m^0 &
\\[1em]
g_{am} \sqrt{n_m^0 n_a^0} &
g_{am} \sqrt{n_m^0 n_a^0} - 2 \alpha \sqrt{n_a^0} &
g_m n_m^0 &
- \omega_n + \xi_m + 2g_m n_m^0 + g_{am} n_a^0 \hspace*{4mm} &
\end{pmatrix}
\end{align}
\end{widetext}
where $\omega_n$ are bosonic Matsubara frequencies, and for brevity we have introduced
\begin{align}
\xi_a \equiv \frac{k^2}{2m} - \mu
\qquad
\xi_m =\frac{k^2}{4m} - 2\mu + \nu
\end{align}
The condition $\det(G_0^{-1})=0$ gives the excitation spectrum.
Notice that that the $G_0^{-1}$ matrix still depends on the chemical potential $\mu$: when we substitute the mean field value of $\mu$ found from Eq.\eqref{eq:mu_equil_1} and \eqref{eq:mu_equil_2} we describe the fluctuations around the chosen mean field solution, ignoring the feedback of the fluctuations on the chemical potential itself.
The dispersion of the gapless mode is convex in both phases and thus allows spontaneous phonon emission. The lifetime is known at large $\nu$ \cite{beliaev1958,vincentliu1997}, either positive or negative, where the gapped mode is inaccessible and the system reduces to a single component BEC.
Close to the transition the picture is more complicated, as the phonon and gapped modes are coupled: in this regime, the low energy behavior of the system is well described by an effective theory of two real fields \cite{radzihovsky2008,lee2004}.

\hypertarget{spectrum-in-msf-phase}{%
\subsection{Spectrum in MSF phase}\label{spectrum-in-msf-phase}}
In the MSF phase the gapped excitation is protected by $\mathbb{Z}_2$ symmetry, so the only allowed process at low momentum it the emission of a phonon. The kinematics of this decay are such that the line is sharp until the group velocity matches the speed of sound at the critical momentum $k_c$, then we expect a decay rate proportional to $(k-k_c)^3$, based on general considerations \cite{abrikosov1963} for phonons.
There are two modes: one particle--like excitation $E_a$ with odd $\mathbb{Z}_2$ charge and one molecule like $E_m$ with even charge:
\begin{align}
\label{eq:msf_Em} E_m &= \sqrt{\epsilon _m \left(2 g_m n_m+\epsilon _m\right)}
&\mathbb{Z}_2 \ \text{even}
\\
\label{eq:msf_Ea} 
E_a &= \sqrt{
(\epsilon_a - \epsilon^+)
(\epsilon_a - \epsilon^-)
}
&\mathbb{Z}_2 \ \text{odd}
\end{align}
In these expressions the free atomic and molecular dispersions are 
\begin{align}
\epsilon_a = \frac{k^2}{2m}
\qquad
\epsilon_m = \frac{k^2}{4m}
\end{align}
and we have defined
\begin{align}
    \epsilon^\pm = \frac{\nu}{2} +  \left(\frac{g_m}{2}  - g_{\text{am}}
    \right) n_m \pm 2 \alpha  \sqrt{n_m}
\end{align}
The dispersion relations $E_{a,m}$ are plotted in the left inset of Fig.\ref{fig:phase_diagram_finitealpha}. We note that Eqs.\eqref{eq:mu_equil_1} and \eqref{eq:mu_equil_2} have already been used to simplify these expressions. The molecular branch $E_m$ is phonon-like; it is the Goldstone mode arising from the breakdown of $U(1)$ to $\mathbb{Z}_2$, with the speed of sound given by the familiar expression for a single component mean field BEC \cite{andersen2004} (the $2m$ in the denominator is the mass of a molecule)
\begin{align}
    \label{eq:msf_speed_sound}
    c_\text{MSF} = \frac{\partial E_m}{\partial k}\bigg|_{k =0} = \sqrt{\frac{g_m  n_m}{2 m}}
\end{align}
The atomic branch $E_a$ is gapped, and is characterized at small momentum by a gap $\Delta_a$ and an effective mass $m_\text{MSF}$
\begin{align}
    E_a \simeq \Delta_a + \frac{k^2}{2 m_\text{MSF}}
\end{align}
as $\nu \to -\infty$ we find
\begin{align}
\Delta_a \simeq
\frac{| \nu | }{2}
+ n_m \left(g_{am}- \frac{g_m}{2}\right)
-\frac{4 \alpha ^2 n_m}{| \nu | }
\end{align}
The leading term $|\nu|/2$ corresponds to breaking a molecule and creating two atomic excitations. The second term has a simple physical interpretation: removing a molecule from the condensate lowers the energy by $n_m g_m/2$ because that molecule does not interact with the others anymore, and it increases the energy by $g_\text{am} n_m$ as now there are two extra atoms interacting with the molecular condensate.
In the same limit, the effective mass $m_\text{MSF}$ of the gapped excitation is 
\begin{align}
    m_\text{MSF} \simeq m \left( 1 - \frac{8 \alpha^2 n_m}{\nu^2} \right)
\end{align}
This reduces to the atomic mass if there is no interconversion or if the binding energy is very large.

\hypertarget{spectrum-in-asf-phase}{%
\subsection{Spectrum in ASF phase}\label{spectrum-in-asf-phase}}
The spectrum in this phase has a closed form expression, but it is very complicated and will not be produced here. The qualitative features are the same as in the MSF phase: there is one gapless mode and one gapped one (see right inset of Fig.\ref{fig:phase_diagram_finitealpha}). The gapless mode is again the Goldstone mode arising from breaking $U(1)$, and we can gain some insight about the gapped one looking at the mean field energy Eq.\eqref{eq:mf_energy}: the Feshbach term is proportional to $\cos(2 \theta_a - \theta_m)$, which gaps the out-of-phase oscillation. The excitations are not protected by parity, which is broken.
Deep in the ASF phase $\nu \to \infty$, the spectrum reduces to \footnote{the subscripts match the notation of Sec.\ref{sub:low_energy_asf_effective_theory}, where we use $\theta$ to denote the phonon and $\phi$ the gapped mode}
\begin{align}
    E_\theta = c_\text{ASF} k + \gamma k^3
\qquad
    E_\phi = \Delta_\phi + \frac{k^2}{2 m_\text{ASF}}
\end{align} 
The parameters in these expressions are, in the $\nu \to \infty$ limit 
\begin{align}
\label{eq:asf_large_nu_parameters_c}
    c_\text{ASF} &=
    \sqrt{\frac{g_a n_a}{m}}
    \left( 1 - \frac{\alpha^2}{\nu g_a}\right)
    \\
\label{eq:asf_large_nu_parameters_gamma}
    \gamma &=
    \frac{1}{8 m c^2}
    \left( 1 + \frac{\alpha^2}{\nu g_a} \right)
    \\
\label{eq:asf_large_nu_parameters_delta}
    \Delta_\phi &= \nu \left( 1 + \frac{(g_{am} - 2g_a)n_a}{\nu} + \frac{8 g_a n_a\alpha^2}{\nu^2 g_a} \right)
    \\
\label{eq:asf_large_nu_parameters_m}
        m_\text{ASF} &= 2 m \left( 1 - \frac{4 n_a\alpha ^2}{\nu^2}\right)
\end{align} 
As $\nu \to \infty$ the speed of sound $c$ reduces to the expression for a weakly interacting BEC, while the gap and the effective mass show that in this limit the excitation is the formation of a molecule. This intuitive understanding breaks down closer to the transition, where the excitations are linear combinations \cite{radzihovsky2008} of all bare atomic and molecular fields $(\psi_a, \bar{\psi}_a, \psi_m, \bar{\psi}_m)$.
The convexity of $E_\theta$ (since $\gamma>0$) implies that finite $k$ phonons can spontaneously emit lower momentum phonons.
The first subleading correction to the gap has like in the MSF phase a simple interpretations: the excitation energy increases by $g_{am} n_a$ because the newly formed molecule interacts with the atoms, but decreases by $2 g_a n_a$ since two atoms that were interacting with the (atomic) condensate disappeared.

\section{Excitation Lifetimes}
\label{sec:lifetimes}
In Sec.\ref{sec:bogoliubov_theory_of_the_molecular_superfluid} we derived the mode structure of the system neglecting the interactions between excitations, which predicts infinitely long lived quasiparticles. This result is far from reality: in both ASF and MSF phases the gapless mode is damped, and the gapped mode may also decay. The field theory predicts finite lifetimes if we include the interactions beyond mean--field; there are three approaches to set up the perturbative calculation: one could diagonalize the matrix in Eq.\eqref{eq:bogoliubov_propagator_matrix}, which makes the bare propagator diagonal, but the interaction terms become much more complicated and even the lowest order correction to the lifetime requires the evaluation of many diagrams. Alternatively, one could keep the matrix in Eq.\eqref{eq:bogoliubov_propagator_matrix} unchanged and work with a non--diagonal propagator, keeping the interaction simple but requiring the evaluation of off--diagonal self energies: we have to solve 
\begin{align}
\det (G_0^{-1}(\omega,k) - \Sigma(\omega,k)) = 0
\end{align}
for $\omega$, where $G_0$ and $\Sigma$ are $4 \times 4$ dense matrices. The last approach, the one we use in this section, is to restrict the range of validity of the theory to low momenta, and write an effective theory with fewer degrees of freedom: in terms of these fields the bare propagator is diagonal, and computing the lifetime to leading order requires the evaluation of a single diagram. 
We use this low energy theory to compute the one-loop decay rate of the gapped modes.

\subsection{MSF Phase}
\label{sub:msf_phase_lifetimes}

\subsubsection{MSF Kinematics} 
\label{ssub:msf_kinematics}
The dispersion relations for the two modes are given in Eq.\eqref{eq:msf_Em} and \eqref{eq:msf_Ea} (see also Fig.\ref{fig:phase_diagram_finitealpha}).
The speed of sound has a simple expression Eq.\eqref{eq:msf_speed_sound},
while the group velocity of the atomic excitation is complicated due to the
square root, but simplifies in the large (negative) $\nu$ limit
\begin{align}
v_a = \frac{\partial E_a}{\partial k} = \frac{k}{m} + \frac{8 \alpha ^2 n_m k}{\nu ^2 m} + O \left( \frac{1}{ \nu^3 }  \right)
\end{align} 
The critical momentum $k_c$ where the speed of sound matches the group velocity of the excitation is then
\begin{align}
\label{eq:k_c}
k_c \simeq
\frac{mc \nu ^2}{\nu ^2+8 \alpha ^2 n_m}
\quad
\text{if} \
|\nu| \gg \left| \frac{4 \alpha^2}{2g_{am} - g_m}\right|
\end{align} 
This momentum scale must be compared to 
\begin{align}
k_m = 2 \sqrt{2m g_m n_m},
\end{align}
the momentum at which the molecular spectrum crosses over from linear to quadratic: taking the ratio 
\begin{align}
\frac{k_c}{k_m} \simeq \frac{1}{2} \left( 1 - \frac{8 \alpha^2 n_m}{ \nu^2 } \right)
\end{align}
shows that $k_m > 2 k_c$, so that when the decay is kinematically allowed the dispersion of the molecular excitation is to a good approximation still linear, and we are justified in considering the speed of sound constant up to $k_c$.

\subsubsection{Low energy MSF theory}
\label{sec:low_energy_msf_effective_theory}

\begin{figure}
    \centering
    \includegraphics[width=0.6\linewidth]{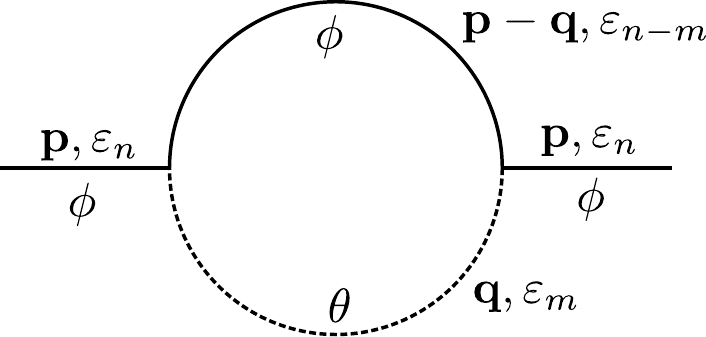}
    \caption{One loop diagram contributing to the imaginary part of the self energy $\Sigma_\phi$ in the MSF phase. In this diagram a solid line denotes the propagator of the massive mode $\phi$, while the dashed line the propagator of the phonon $\theta$.}
    \label{fig:one_loop_msf}
\end{figure}

The first step is to pick the phase for the molecular field so that the vacuum expectation value is real and positive, and reparametrize the fields as
\begin{align}
\Psi_a = \phi + i \chi
\quad
\Psi_m = \sqrt{n_m^0 + \rho} \, e^{i \theta}
\end{align}
where $(\phi,\chi,\rho,\theta)$ are real. Assuming as usual that the fluctuations around the mean field solution are small, we drop from the action terms quartic in $\chi$ and $\rho$, and compute the resulting Gaussian integrals over these two variables \cite{radzihovsky2008}. We also drop terms quartic in $\phi$, as they cannot give a contribution to the imaginary part of the one--loop self energy.
After integration, low energy action is
\begin{align}
S_{\mathrm{eff}}\left[\theta, \phi\right]=S_{\mathrm{SF}}\left[\theta\right]+S_{\mathrm{I}}[\phi]+S_{\mathrm{int}}\left[\theta, \phi\right]
\end{align}
where the three contributions are
\begin{align}
&{S}_{\mathrm{SF}}= \frac{1}{2 g_m}  \int d\tau d \bm{r} \, \left( \left(\partial_{\tau} \theta\right)^{2}+c_\text{MSF}^2 \left(\nabla \theta\right)^{2}\right) 
\\
&{S}_{\text {I}}= \int d\tau d \bm{r} \, \bigg[ \frac{1}{2 \alpha \sqrt
{n_m^0}} \left(\partial_{\tau} \phi\right)^{2}+\frac{(\nabla \phi)^{2}}{2 m}-\mu_{R} \phi^{2} \bigg]
\\
&{S}_{\text{int}}=\frac{\mathrm{i}}{2} \int d\tau d \bm{r} \, \phi^{2} \partial_{\tau} \theta,
\end{align}
The shifted chemical potential is $\mu_R = \mu + 2 \alpha \sqrt{n_m^0}$.
The effective degrees of freedom are a phonon $\theta$ with linear dispersion and a Klein-Gordon particle $\phi$, which matches the structure of the spectrum discussed in Sec.\ref{spectrum-in-msf-phase} and Fig.\ref{fig:phase_diagram_finitealpha}. The form of the vertex is the one we expect for the interaction between a phonon and a massive particle: at $q=0$ there must be no interaction, as a perfectly uniform medium does not scatter.
the $S_\text{int}$ term couples $\phi$ to $\theta$: in this effective theory the leading mechanism for sound absorption is the production of a pair of massive particles, not two phonons like in liquid Helium \cite{bhatt1974}, so that sound absorption measures the decay rate of $\phi$.
In the on--shell approximation, the decay rate of $\phi$ is 
\begin{align}
    \gamma_\text{MSF}(k) = - \text{Im} \Sigma_\phi(E_a(k), k)
\end{align}
and the only diagram contributing to $\Sigma_\phi$ to one--loop order is the one presented in Fig.\ref{fig:one_loop_msf}. Evaluating the diagram at zero temperature we find
\begin{align}
\begin{split}
\gamma_\text{MSF}(k) = 
&\frac{g_m c_\text{MSF}}{64 \pi^2} 
\sqrt{ \frac{\alpha \sqrt{n_m^0}}{|\mu_R|} }
\\
&\times \int d\mathbf{q} \, \delta 
\big(
E_a(k) - c_\text{MSF} q - E_a(\mathbf{k} - \mathbf{q})
\big)
\end{split}
\end{align}

\begin{figure}
    \centering
    \includegraphics[width=\columnwidth]{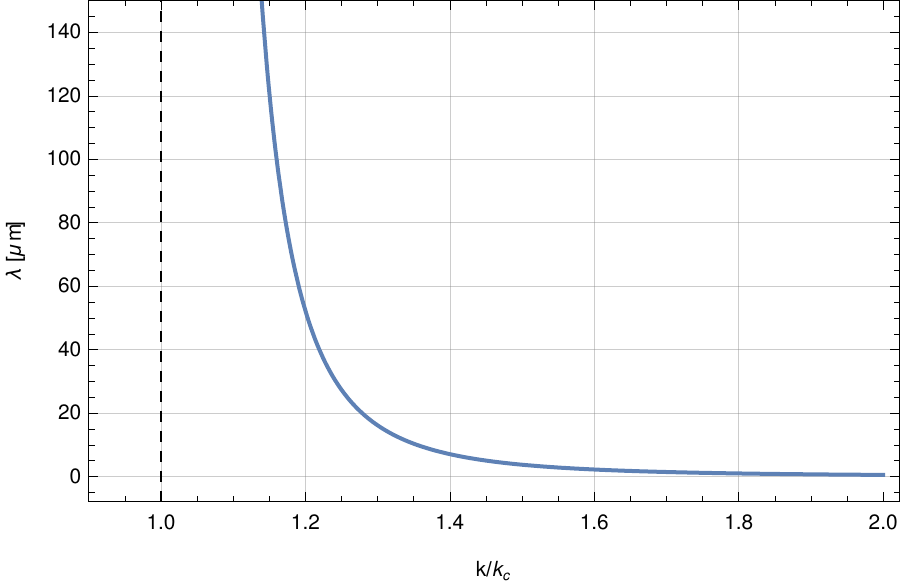}
    \caption{Predicted decay length $\lambda$ of the gapped mode in MSF phase, as a function of momentum (in units of the critical momentum $k_c$) using data for the $47.97$G Feshbach resonance of Cesium \cite{chin2010}, assuming a condensate density $n_0 = 10^{14} \text{cm}^{-3}$ and a constant magnetic field $B = 47.87$G. The dashed vertical line is the asymptote at $k = k_c$, where the decay length diverges.}
    \label{fig:decay_length}
\end{figure}
We evaluate the integral in this expression using the technique described in the Appendix, which finally gives
%
\begin{align}
\label{eq:gamma_msf}
    \gamma_\text{MSF}(k) = \frac{m c_\text{MSF}^2}{12 \pi} \frac{(k-k_c)^3}{n_m^0} \sqrt{\frac{\alpha \sqrt{n_m^0}}{|\mu_R|}}
\end{align}
This result is compatible with the well--known $k^3$ dependence for the lifetime at the endpoint of a spectrum due to soft phonon emission (see Ref.\cite{abrikosov1963}, Sec.(26.3)), and has a few simple experimental consequences: the decay rates are inputs to the calculation of spectral functions, which are directly observable in some experimental setups (for example, see Refs.\cite{valdes-curiel2017,petter2019,allard2016}). Furthermore, the existence of the threshold momentum $k_c$ in the MSF phase (see Fig.\ref{fig:phase_diagram_finitealpha}, left inset) implies that it behaves as a ``momentum funnel'' in transmission experiments: neglecting reflection at the phase boundary, an atom propagating with momentum $k$ through a cloud of MSF bosons will not lose any energy if $k < k_c$, but emit phonons and decrease its momentum until it reaches $k$ if initially $k > k_c$. To see this effect the cloud has to be thick compared to the absorption length $\lambda$, defined as
\begin{align}
\lambda(k) = \frac{\hbar v_\text{MSF}(k)}{\gamma_\text{MSF}(k)}
\end{align}
where $v_\text{MSF} = \partial_k E_a(k)$ is the group velocity of the gapped excitation: if the atom cloud is thinner than $\lambda(k)$ the atom leaves the cloud before emitting phonons. To get an idea of the typical size of $\lambda$ see Fig.\ref{fig:decay_length}, where give an example using data for one of the Feshbach resonances in cesium. Notice that $\lambda$ is sensitive to the condensate density $n_m^0$ --a higher density makes the cloud more opaque-- and the background magnetic field, through the shifted chemical potential $\mu_R$ (see Eq.\eqref{eq:gamma_msf}). We also have to keep into account the trapping time: the cloud must remain trapped while the projectile traverses it. In particular, to see the sharp feature corresponding to $k_c$, the cloud must survive for a time
\begin{align}
  \Delta t \geq \frac{L}{c}
\end{align}
where $L$ is the diameter of the sample. Using the same parameters as in Fig.\ref{fig:decay_length} for the speed of sound and assuming $L \simeq 20 \mu$m we must have $\Delta t \geq 5$ms, which is compatible with the current experimental possibilities \cite{zhang2021}.

\subsection{Lifetime of the gapless mode}
For completeness we report the analogous result for the decay of one phonon into two phonons. When the binding energy is very large the phonon decay rate is the same as in the single component condensate, and it was first derived by Belyaev \cite{beliaev1958} at zero temperature. The decay rate does not have an elementary expression. In the small and large momentum limits it reduces to
\begin{align}
\label{eq:gamma_single_component}
\gamma_{T=0}(\mathbf{k})=
\begin{cases}
\dfrac{3 k^{5}}{640 \pi (2m) n_m^{0}} & k \ll k_c
\\[3mm]
  \dfrac{4 \pi a^2 n_m^0 k}{2 m} & k \gg k_c
\end{cases}
\end{align}
where $a$ is the s--wave scattering length. We expect this same behavior for the phonon in both phases, as long as $c k \ll \Delta$, where $c$ is the speed of sound in the phase and $\Delta$ the gap. For a discussion of the damping properties of the gapless mode in the critical regime see Ref.\cite{lee2004}.

\hypertarget{cherenkov-condition-in-asf-phase}{%
\subsection{ASF phase}\label{cherenkov-condition-in-asf-phase}}

\subsubsection{ASF kinematics} 
\label{ssub:asf_kinematics}
In the ASF phase the phonons can spontaneously emit other phonons, and so can the gapped mode: there is no symmetry protecting the decay. At large momentum there are two processes available for the decay of a gapped excitation: pair production of two massive particles, and phonon emission. We focus on the long wavelength physics, and only discuss phonon emission. 
Deep in the ASF phase at $k=0$ a gapped excitation emits two back--to--back phonons with momentum $k_0$, fixed by energy conservation 
\begin{align}
\begin{split}
\Delta_\phi &= 2 c_\text{ASF} k_0 \\ &\rightarrow k_0 \simeq \frac{\nu }{2 c_\text{ASF}} + \frac{\alpha ^2+g_a n_a \left(g_{\text{am}}-2 g_a\right)}{2 c_\text{ASF} g_a}
\end{split}
\end{align}
and at finite $k$ the emission of two back--to--back phonons is
allowed if 
\begin{align}
\begin{split}
\Delta_\phi + \frac{k^2}{2 m_{\text{ASF}}} &= c_\text{ASF} q + c_\text{ASF} |\bm{k} - \bm{q}| 
\\
&= c_\text{ASF}(k + 2q)
\end{split}
\end{align}
assuming $\bm{q} = -q \bm{\hat{k}}$ where $\bm{\hat{k}}$ is the unit vector in the direction of $\bm{k}$.
Requiring $q>0$ we get
\begin{align}
k^2 - 2 m_\text{ASF} c_\text{ASF} k + 2 m_\text{ASF}\Delta_\phi > 0
\end{align}
which is true as long as $m_\text{ASF} c_\text{ASF}^2 < 2 \Delta_\phi$. Working at leading order in $\nu$ and using Eqs.\eqref{eq:asf_large_nu_parameters_c}-\eqref{eq:asf_large_nu_parameters_m} this condition reduces to
\begin{align}
    g_a n_a < 2 \nu 
\end{align} 
and holds if the density is low enough. 

\subsubsection{Low energy ASF theory} 
\label{sub:low_energy_asf_effective_theory}
\begin{figure}[h]
    \centering
    \includegraphics[width=0.6\linewidth]{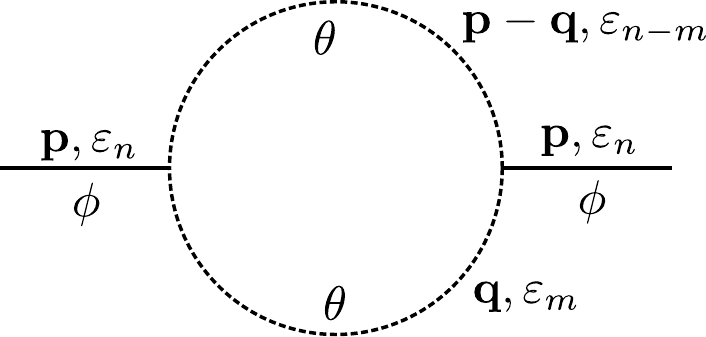}
    \caption{One loop diagram contributing to the imaginary part of the self energy in the ASF phase. In this diagram a solid line denotes the propagator of the massive mode $\phi$, while the dashed line the propagator of the phonon $\theta$.}
    \label{fig:one_loop_asf}
\end{figure}

In the ASF phase both the molecules and the atoms are condensed. The Feshbach terms locks the phases of the atomic and molecular order parameters, so it is convenient to redefine the atomic fields as
\begin{align}
  \Psi_a \to e^{i \theta_m/2} \Psi_a 
\end{align}
where $\theta_m$ is the phase of the molecular field:
\begin{equation}
  \Psi_m = \sqrt{n_m^0 + \rho_m} e^{i \theta_m}
\end{equation}
We choose the phase of the molecular field so that in equilibrium $\theta_m = 0$. Then for small fluctuations around the ASF equilibrium condition
\begin{align}
  \Psi_a = \sqrt{n_a^0} + \psi_a
\end{align}
and we expand to second order in $\psi_a$. The terms linear in $\psi_a$ drop out. We further decompose the atomic fluctuations in their real and imaginary parts:
\begin{align}
\psi_a = \phi + i \chi
\end{align}
As pointed out in Ref.\cite{radzihovsky2008}, the fluctuations in the $\phi$ direction reach criticality before the ones in $\chi$, so we drop terms of third and higher in $\chi$ and perform the resulting $\chi$ Gaussian integral. Deep in the ASF phase the $\phi$ fluctuations are also small, and integrating over $\phi$ first will give the same results.
After integration the action is 
\begin{widetext}
\begin{align}
\begin{split}
\label{eq:asf_action}
S = \int d\tau d\bm{r} 
\bigg[ 
\frac{(\partial_\tau\theta)^2}{2 g_m} + \frac{n_m^0}{4m} (\grad \theta)^2 
+ i \sqrt{n_a^0} (1 - y) \phi \partial_\tau \theta 
+\frac{ (\partial_\tau \phi)^2}{2\alpha \sqrt{n_m^0}}
+ \phi \bigg( -\frac{\nabla^2}{2m} - \tilde{\mu} \bigg) \phi
\\
+ \bigg(2 g_a - \frac{g_m}{2} y^2\bigg) \sqrt{n_a^0} \phi^3
+ \frac{i}{2} (1 - y) \phi^2 \partial_\tau \theta  - y \sqrt{n_a^0} \phi \frac{(\nabla \theta)^2}{4m}
\bigg]
\end{split}
\end{align}
\end{widetext}
Where we have relabeled $\theta_m \to \theta$, and we have dropped terms that cannot produce one loop diagrams. We have also defined
\begin{align}
y &= \frac{2}{g_m} \bigg( g_{am} - \frac{\alpha}{\sqrt{n_m^0}} \bigg)
\\
\tilde{\mu} &= \mu - 3 g_a n_a^0 + g_{am} n_m^0 - 2 \alpha \sqrt{n_m^0} - \frac{g_m}{2} y^2 n_a^0
\end{align}
The low energy description of the ASF phase is an interacting theory of a phonon $\theta$ and a gapped Klein-Gordon field $\phi$. 
Unlike in the MSF phase, the decay of one gapped excitation into two phonons is allowed here, and is in fact the only allowed process at low enough momentum. 
Furthermore, at very small momentum the there is no phase space for the process $\phi \to \phi + \theta$, and energy conservation forbids $\phi \to \phi + \phi$, so we only keep the $\phi (\nabla \theta)^2$ vertex. 
The term $i \sqrt{n_a^0} \phi \partial_\tau \theta$ mixes the two fields, but it does not gap the phonon and is safe to neglect in a first approximation, as its main effect is to give a correction to the gap and the speed of sound. This approximation simplifies the calculation of the decay rate of the gapped mode: in the on--shell approximation
\begin{align}
    \gamma_\text{ASF}(k) = - \text{Im} \Sigma_\phi(E_\phi(k),k)
\end{align}
We compute the self energy $\Sigma_\phi$ to one loop, evaluating the diagram in Fig.\ref{fig:one_loop_asf} at zero temperature.
The decay rate of the gapped excitation is then
\begin{widetext}
\begin{align}
\gamma_\text{ASF}(k) = \frac{Z_\theta^2 \lambda^2}{32 \pi c_\text{ASF}^2} 
\int d \mathbf{q} \,
\frac{(\mathbf{q} \cdot (\mathbf{k} - \mathbf{q}))^2}{q |\mathbf{k} - \mathbf{q}|}
\delta \big(E_\phi(k) - c_\text{ASF} q - c_\text{ASF} |\mathbf{k} - \mathbf{q}| \big)
\end{align}
\end{widetext}
where we have defined the parameters
\begin{align}
Z_\theta \simeq g_m + \frac{\alpha n_a^0}{4 (n_m^0)^{3/2}}
\qquad
\lambda \simeq \frac{y \sqrt{n_a^0}}{2m}
\end{align}
We evaluate the integral using the method described in the Appendix, which finally gives
\begin{align}
\label{eq:gamma_asf}
\gamma_\text{ASF}(k) = \frac{Z_\theta^2 \lambda^2}{32 \pi c_\text{ASF}^2} \bigg( \frac{E_\phi(k)^4}{16 c_\text{ASF}^5} - \frac{E_\phi(k)^2 k^2}{c_\text{ASF}^3} + o(k^2) \bigg)
\end{align}
The parametric dependence of $\gamma_\text{ASF}$ is intuitively clear: when the gap is larger there is more phase space available, which leads to faster decay, while for larger speed of sound there is less phase space and the decay slows down. The dependence on $E_\phi^4$ is the same reported in \cite{bhatt1974}, but in that work the particle disintegrating is itself a phonon, not a gapped excitation. This result implies that the gapped mode is strongly damped, and will be hard to see in experiment.

\subsection{Effective theory range of validity} 
\label{sub:effective_theory_range_of_validity}
For our approximation to hold, we require the fluctuations of the $\phi_I$ field, which we integrated out, to remain stable while the $\phi_R$ field undergoes spontaneous symmetry breaking. Using the mean field value of the chemical potential 
\begin{equation}
    \mu = g_a n_a^0 + g_{am} n_m^0 - 2 \alpha \sqrt{n_m^0}
\end{equation}
the validity condition for the ASF low energy theory is
\begin{equation}
    \left( g_{am} - \frac{\alpha}{\sqrt{n_m^0}} \right)^2 > g_a g_m
\end{equation}
which is verified if $n_m^0$ is small enough.


\section{Conclusions and Outlook} 
\label{sec:conclusions}

In this work we studied the damping of  excitation modes in the ASF and MSF phases of a single-component ultracold atomic gas. 
In both phases, away from the critical regime, the phonon damps according to the Belyaev theory \cite{beliaev1958}, as expected of a single-component Goldstone mode in a non-relativistic theory. 
In the ASF phase the gapped mode has a finite lifetime even in the zero momentum limit, while in the MSF phase we find a Cherenkov threshold: below the critical momentum $k_c$ there is no damping, while for $k>k_c$ the dissipation rate is proportional to~$(k-k_c)^3$.
We discussed how this property of the MSF phase should be observable in experiments in which an atom is transmitted through an MSF cloud: the cloud acts as a momentum funnel, suppressing the transmission amplitude at momenta larger than the threshold and shifting the distribution of transmitted atoms to lower energy. 

Such transmission spectroscopy experiments promise to provide great insight into the properties of the MSF and other more exotic ultracold phases.
Our work serves as a building block to constructing a complete dissipative transport theory for such near--equilibrium dynamics in the MSF and ASF phases. 
It would be interesting to understand how our results are modified as the system approaches the critical regime, where the assumption of small fluctuations around the mean-field breaks down. 
Another promising research direction is the study of heteronuclear condensates \cite{lam2022,thalhammer2009,warner2021,grobner2016}, for which strictly speaking our theory does not apply --already at mean field level the phase diagram changes. 
Nonetheless, kinematic arguments still suggest \textbf{}the existence of a Cherenkov threshold, and a damping structure qualitatively similar to the homonuclear case.


\begin{acknowledgements}
The authors are grateful to A. Chandran, L. Radzhihovsky, and C. Chamon for stimulating discussions. 
This work was supported by the National Science Foundation through the awards PHY-1752727 (C.R.L.) and DMR-1906325 (G.G.).
\end{acknowledgements}

\appendix

\section{Off--shell Cherenkov}
\label{sub:off_shell_cherenkov}
In this section we catalog some useful tricks for the kind of
integration that shows up when dealing with one loop calculations.
Consider a massive particle $\phi$ emitting a phonon $\gamma$:
\begin{align}
E_\phi(p) = \Delta + \frac{p^2}{2m}
\qquad
E_\gamma(p) = c p
\end{align} We compute the phase space for this process: the process of a particle with momentum $p$ and energy $\epsilon$ spontaneously emitting a phonon has phase space
\begin{align}
g(p,\epsilon) = \int \frac{d\mathbf{q}\, d\omega}{(2\pi)^4}\, \delta(\omega - c q) \delta\bigg(\epsilon - \omega - \frac{(\mathbf{p}-\mathbf{q})^2}{2m}\bigg)
\end{align} where we allow $\epsilon$ to take any value: we include processes
where the emitted $\phi$ is virtual, as is the case in one loop
perturbation theory. When evaluating a diagram of the kind discussed in Sec.\ref{sec:lifetimes} the only difference is some extra factor of $q$ in the integrand, the nontrivial part is dealing with the energy--conserving Dirac delta .To evaluate this integral it is convenient to measure the energy
$\epsilon$ with respect to the threshold energy
$\epsilon_\text{th}$, defined as the lowest energy where the decay is
allowed, starting with a particle with momentum $p$ and energy
$\epsilon = E_\phi(p)$: the threshold energy is then the minimum as a
function of $q$ and $\Omega$ of \begin{align}
E = \Delta + \frac{p^2 + q^2 - 2 p q \Omega}{2m} + cq
\end{align} on the minimization domain $q \in [0,\infty), \Omega \in [-1,1]$.
The derivatives of $E$ are \begin{align}
\frac{\partial E}{\partial \Omega} = - \frac{pq}{m} < 0
\qquad
\frac{\partial E}{\partial q} = \frac{q - p \Omega + mc}{m}
\end{align} The derivative of $E$ with respect to $\Omega$ is always
negative, the function assumes its minimum on the upper edge of the
domain: the optimal value is $\Omega^*=1$ (at threshold the phonon is
emitted forward). For the other equation we have to discuss two cases
separately:
\begin{itemize}
    \item If $p < mc$ then we always have $\partial_q E >0$,
thus the optimal $q$ and threshold energy are
\begin{align}q^* = 0 \qquad \epsilon_\text{th} = \Delta + \frac{p^2}{2m}\end{align} which is
consistent with the discussion above: when $p<mc$ the on shell process
cannot happen (it is only allowed to emit a zero momentum phonon, but
there is no such thing). Later we study the off-shell case, which is
less trivial.
\item If $p < mc$ the minimum is at nontrivial $q$, and
gives
\begin{align}q^* = p - mc \qquad \epsilon_\text{th} = \Delta + pc - \frac{mc^2}{2}\end{align}
In the next section we will derive the phase space in these two cases.
The space space is $0$ when $\epsilon < \epsilon_\text{th}$ so the
formulae we derive should be understood as holding only for
$\epsilon > \epsilon_\text{th}$.
\end{itemize}

\hypertarget{phase-space-for-p-mc}{%
\subsection{\texorpdfstring{Phase space for
$p < mc$}{Phase space for p \textless{} mc}}\label{phase-space-for-p-mc}}

The integral over $\omega$ is trivial: \begin{align}
g(p,\epsilon) =
\int \frac{d \mathbf{q}}{(2\pi)^4}
\delta \bigg(
\epsilon - cq
-\frac{p^2}{2m}
-\frac{q^2}{2m}
+ \frac{pq\Omega}{m}
- \Delta
\bigg)
\end{align} use the threshold energy to eliminate $\Delta$ and define
$\Delta \epsilon = \epsilon - \epsilon_\text{th}$: \begin{align}
g(p,\epsilon) =
\int \frac{d \mathbf{q}}{(2\pi)^4}
\delta \bigg(
\Delta\epsilon - cq
-\frac{q^2}{2m}
+ \frac{pq\Omega}{m}
\bigg)
\end{align} at this point we have two options: integrate over $\Omega$ or
integrate over $q$. They are equivalent of course, but it is useful to
show the details to understand how the two approaches differ.

Let's integrate over $\Omega$ first. The support and Jacobian factor of the Dirac delta are \begin{align}
\Omega_0 =
\frac{m}{pq}
\bigg(
\frac{q^2}{2m} + cq - \Delta \epsilon
\bigg)
\qquad
\frac{1}{|f'(\Omega_0)|} = \frac{m}{pq}
\end{align} where $f(\Omega)$ is the argument of the Dirac delta in the
integrand. After integrating over $\Omega$ we get \begin{align}
g(p,\epsilon) =
\int \frac{d q}{(2\pi)^3} q^2 \frac{m}{pq}
\bigg[
-1 <
\frac{m}{pq}
\bigg(
\frac{q^2}{2m} + cq - \Delta \epsilon
\bigg)
< 1
\bigg]
\end{align} where the square brackets represent the
Heaviside theta ([condition] evaluates to $1$ when the condition
is true and $0$ otherwise). The restriction on the $q$ domain is equivalent to
\begin{align}
\begin{split}
&\bigg[
-1 <
\frac{m}{pq}
\bigg(
\frac{q^2}{2m} + cq - \Delta \epsilon
\bigg)
< 1
\bigg]
=
\\
&\bigg[
\frac{q^2}{2m} + cq - \Delta \epsilon > -\frac{pq}{m}
\bigg]
\bigg[
\frac{q^2}{2m} + cq - \Delta \epsilon < \frac{pq}{m}
\bigg]
\end{split}
\end{align}
The first condition is equivalent to
\begin{align}
q > - p - mc + \sqrt{(p+mc)^2 + 2m \Delta\epsilon}
\end{align} (notice the right hand side is always positive), and the second to \begin{align}
\begin{split}
q > p - mc - \sqrt{(p-mc)^2 + 2m \Delta\epsilon}
\\
\text{and} \ q <
p - mc + \sqrt{(p-mc)^2 + 2m \Delta\epsilon}
\end{split}
\end{align} The quantity in the right hand side of the first line is negative, so that condition is automatically verified by $q>0$. The second condition is complicated, but
if we are only interested in the small $\Delta \epsilon$ limit we expand to first order the square root, and find
\begin{align}
\begin{split}
\bigg[
-1 <
\frac{m}{pq}
\bigg(
\frac{q^2}{2m} + cq - \Delta \epsilon
\bigg)
< 1
\bigg]
=
\\
\bigg[
\frac{m \Delta \epsilon}{mc + p}
< q <
\frac{m \Delta \epsilon}{mc - p}
\bigg]
\end{split}
\end{align} so that finally the integral gives
\begin{align}
\begin{split}
g(p,\epsilon) &=
\frac{m}{2p} \frac{(m \Delta \epsilon)^2}{(2\pi)^3}
\bigg(
\frac{1}{(mc - p)^2}
-
\frac{1}{(mc + p)^2}
\bigg)
\\
&= \frac{m^4 c}{4\pi^3} \frac{(\Delta\epsilon)^2}{((mc)^2 - p^2)^2}
\end{split}
\end{align}

If we start by integrating over $q$ instead, the support and Jacobian factor of the Dirac delta are \begin{align}
q_0 = p\Omega - mc + \sqrt{(p\Omega - mc)^2 + 2 m \Delta \epsilon}
\\
\frac{1}{|f'(q_0)|} = \frac{m}{\sqrt{(p\Omega - mc)^2 + 2 m \Delta \epsilon}}
\end{align} so after the integration over $q$ we have \begin{align}
g(p,\epsilon) =
\frac{1}{(2\pi)^3}
\int_{-1}^1 d\Omega \,
\frac{mq_0^2 }{\sqrt{(p\Omega - mc)^2 + 2 m \Delta \epsilon}}
\end{align} The advantage with respect to the previous approach is that there is
no condition restricting the integration domain, but the integrand is
more complicated. It's convenient to change variable to
$x = mc - p\Omega$ (notice that $x>0$) which turns the integral into
\begin{align}
g(p,\epsilon) =
\frac{m}{(2\pi)^3p}
\int_{mc - p}^{mc + p} dx \,
\frac{(-x + \sqrt{x^2 + 2m\Delta\epsilon})^2}{\sqrt{x^2 + 2 m \Delta \epsilon}}
\end{align} expanding in small $\Delta \epsilon$ again, this integral is
approximately \begin{align}
\begin{split}
g(p,\epsilon) &=
\frac{m}{(2\pi)^3p}
\int_{mc - p}^{mc + p} dx \,
x \bigg( \frac{m\Delta\epsilon}{x^2} \bigg)^2
\\&=
\frac{m^3 (\Delta\epsilon)^2}{2(2\pi)^3p}
\bigg(
\frac{1}{(mc - p)^2}
-
\frac{1}{(mc + p)^2}
\bigg)
\\&=
\frac{m^4 c}{4\pi^3} \frac{(\Delta\epsilon)^2}{((mc)^2 - p^2)^2}
\end{split}
\end{align} as expected from the previous section.

In conclusion, the trade-off we make
integrating over $q$ first is that the result integral is harder, but
we don't have to think as hard about inequalities that restrict the
domain: the $\delta(q - q_0)$ is guaranteed to have the support in the
integration interval $[0,\infty)$ as long as we use the positive root
$q_0$.

\hypertarget{phase-space-for-p-mc-1}{%
\subsection{\texorpdfstring{Phase space for
$p > mc$}{Phase space for p \textgreater{} mc}}\label{phase-space-for-p-mc-1}}

The starting point is the same as in the previous section \begin{align}
g(p,\epsilon) =
\int \frac{d \mathbf{q}}{(2\pi)^4}
\delta \bigg(
\epsilon - cq
-\frac{p^2}{2m}
-\frac{q^2}{2m}
+ \frac{pq\Omega}{m}
- \Delta
\bigg)
\end{align} but the momentum of the phonon at threshold $q^*$ and the threshold
energy are different: (we still have $\Omega^* = 1$) \begin{align}
q^* = p - mc \qquad \epsilon_\text{th} = \Delta + pc - \frac{mc^2}{2}
\end{align} Eliminating $\Delta$ using $\epsilon_\text{th}$ we get (as usual
$\Delta \epsilon \equiv \epsilon - \epsilon_\text{th}$) \begin{align}
\begin{split}
g(p,\epsilon) =
\int \frac{d \mathbf{q}}{(2\pi)^4}
\delta \bigg(
&\Delta\epsilon
- c (q - q^*)
\\ &-\frac{(\mathbf{p}-\mathbf{q})^2}{2m}
+\frac{(\mathbf{p}-\mathbf{q}^*)^2}{2m}
\bigg)
\end{split}
\end{align} where $\mathbf{q}^*$ is a vector in the same direction as
$\mathbf{p}$ of length $q^*$. The argument of the Dirac delta
simplifies a little defining $\delta q \equiv q - q^*$: \begin{align}
\begin{split}
g(p,\epsilon) =
\int \frac{d \mathbf{q}}{(2\pi)^4}
\delta \bigg(
&\Delta\epsilon
- c \delta q
-\frac{\delta q^2}{2m}
\\&+ \frac{p}{m} (\Omega - 1) (\delta q + q^*) + c \delta q
\bigg)
\\
=
\int \frac{d \mathbf{q}}{(2\pi)^4}
\delta \bigg(
&\Delta\epsilon
-\frac{(q - q^*)^2}{2m}
+ \frac{pq}{m} (\Omega - 1)
\bigg)
\end{split}
\end{align} Notice that the term quadratic in $q$ is $(q - q^*)^2$, which is
\emph{not} a vector difference, it's the difference of two real positive
numbers. Again we have two possibilities, integrating over $\Omega$ or
$q$ first.

Integrating over $\Omega$ first, the support and Jacobian factor of the Dirac delta are \begin{align}
\Omega_0 = 1 + \frac{m}{pq} \left(\frac{(q - q^*)^2}{2m} - \Delta \epsilon \right)
\qquad
\frac{1}{|f'(\Omega_0)|} = \frac{m}{pq}
\end{align} so integrating over $\Omega$ we get \begin{align}
g(p,\epsilon) =
\int_0^\infty \frac{d {q}}{(2\pi)^3} q^2
\frac{m}{pq}
\big[
-1 < \Omega_0 < 1
\big]
\end{align} The condition restricting the integration domain is equivalent to \begin{align}
\begin{split}
\big[
-1 < \Omega_0 < 1
\big]
=
\bigg[
-\frac{2pq}{m} <
\frac{(q - q^*)^2}{2m} - \Delta \epsilon
\bigg]
\\ \times \bigg[
\frac{(q - q^*)^2}{2m}  < \Delta\epsilon
\bigg]
\end{split}
\end{align} If $\Delta\epsilon$ is small enough, the first condition is always
satisfied, so we only have to consider the second inequality \begin{align}
\begin{split}
[(q - q^*)^2  < 2m \Delta\epsilon]
=
\bigg[q > q^* - \sqrt{2 m \Delta\epsilon}\bigg]
\\ \times\bigg[q < q^* + \sqrt{2 m \Delta\epsilon}\bigg]
\end{split}
\end{align} Notice that $q^* - \sqrt{2m\Delta\epsilon}$ is a strictly positive
number for $\Delta\epsilon$ small enough, so the first condition on
the right hand side restricts the integration domain more than just
asking $q > 0$. The integral is then \begin{align}
\begin{split}
g(p,\epsilon) =&
\frac{m}{p(2\pi)^3}
\int_{q^* - \sqrt{2m\Delta\epsilon}}^{q^* + \sqrt{2m\Delta\epsilon}}
d {q} \, q
\\
=&
\frac{m}{2p(2\pi)^3}
\bigg(
\big(q^* + \sqrt{2m\Delta\epsilon}\big)^2
-
\big(q^* - \sqrt{2m\Delta\epsilon}\big)^2
\bigg)
\\
=&
\frac{m(p - mc)}{p(2\pi)^3} \sqrt{2m\Delta\epsilon}
\end{split}
\end{align}

Integrating over $q$ first, the Dirac delta support and Jacobian are
\begin{align}
\begin{split}
q_0^\pm = q^* + &p(\Omega - 1)
\\ &\pm
\sqrt{(q^* + p(\Omega - 1))^2 - (q^*)^2 + 2 m\Delta\epsilon}
\end{split}
\end{align}
\begin{align}
\frac{1}{|f'(q_0^\pm)|} &=
\frac{m}{ \sqrt{(q^* + p(\Omega - 1))^2 - (q^*)^2 + 2 m\Delta\epsilon} }
\end{align} using $q^* = p - mc$ and plugging in the phase space integral we
have \begin{align}g(p,\epsilon) =
m
\int_{-1}^1 \frac{d\Omega}{(2\pi)^3}
    \frac{(q_0^+ [q_0^+ >0])^2 + (q_0^- [q_0^- >0])^2}{ \sqrt{(mc + p\Omega)^2 - (p - mc)^2 + 2 m\Delta\epsilon} }
\end{align} and we should also require that the argument in the square root
appearing in $q_0^\pm$ is positive, which is a condition on
$\Omega$. Dealing with these inequalities is complicated, 
in this case integrating over $\Omega$ first is the
the best approach.

\bibliography{msf}

\end{document}